# Quantum tunneling of magnetization in lanthanide single-molecule magnets, bis(phthalocyaninato)terbium and bis(phthalocyaninato)dysprosium anions**

*Naoto Ishikawa,[*] Miki Sugita and Wolfgang Wernsdorfer[*]*

Single-molecule magnets (SMMs) are the class of high-spin molecules that exhibit magnetization hysteresis at low temperature, that is the property of macroscopic

[*]  Prof. Dr. N. Ishikawa
     Department of Applied Chemistry, Faculty of Science and Engineering,
     Chuo University, 1-13-27 Kasuga, Bunkyo-ku, Tokyo 112-8551, (Japan)
     E-mail: ishikawa@chem.titech.ac.jp

     M. Sugita
     Department of Chemistry, Tokyo Institute of Technology
     O-okayama, Meguro-ku, Tokyo 152-8551, (Japan)

     Dr. W. Wernsdorfer,
     Laboratoire Louis Néel
     CNRS, BP 166
     25 Avenue des Martyrs, 38042 Grenoble Cedex 9 (France)
     Fax: (+33) 4-7688-1191
     E-mail: wernsdor@grenoble.cnrs.fr

[**] This work was partly supported by the EC-TMR Network "QuEMolNa" (MRTN-CT-2003-504880) and by Grant-in-Aid for Science Research No. 15550046 from the Ministry of Education, Science Sports and Culture of Japan.

magnets.[1-5] Most SMMs are composed of several transition-metal ions, whose spins are coupled by strong exchange interactions to give a large effective spin with a predominant uniaxial anisotropy. The quantum nature of SMMs is manifested by staircase hysteresis loops[5-6], temperature independent relaxation[7-12], and quantum phase interference.[13] The discovery of these phenomena led to potential applications for quantum computing.[14]

The finding of slow magnetization relaxation in lanthanide complexes has opened the possibility of constructing SMMs containing only a single metal ion as a magnetic center.[15] Alternating current (ac) magnetic susceptibility measurements have been reported for bis(phthalocyaninato)terbium anion [$Pc_2Tb$]$^-$ (Pc = dianion of phthalocyanine) and an isostructural dysprosium complex [$Pc_2Dy$]$^-$ above 2 K.[15,16] One of the important results was that slow magnetization relaxation was observed in the temperature ranges significantly higher than those of previously known transition-metal SMMs. For example, the peak positions in the $\chi''$ vs. $T$ plot ($\chi_M''$ refers to as the out-of-phase component of ac susceptibility) of the Tb and Dy complexes are at 40 K and 10K, respectively, for 1 kHz ac-frequency.[15,16] The Arrhenius analysis showed that the dominant relaxation path in the high temperature range, above 25 K for [$Pc_2Tb$]$^-$ and 3 K for [$Pc_2Dy$]$^-$, is a thermally activated Orbach process involving excited substates in the ground multiplet.[16] Hysteresis loops were measured at 1.7 K, however, no clear evidence of quantum tunneling was observed because of instrumental limitations.[16]

In this communication, we report magnetic hysteresis measurements for these lanthanide single-ion SMMs in the subkelvin temperature range. Clear evidence of quantum tunneling of magnetization (QTM) is presented for the first time. We found that the quantum process in these single-ion SMMs is due to the resonant quantum tunneling between entangled states of electron and nuclear spin systems, which has been first reported for a scheelite-structured compound LiYF$_4$ doped with trivalent Ho ions.[17]

The compounds were prepared as reported in literature[18,19] with certain modifications.[20] The doped single crystalline samples were prepared by recrystallization from a mixed solution of [Pc$_2$Ln]$^-$·TBA$^+$ (Ln = Tb or Dy; TBA$^+$ = tetrabutylammonium cation) and [Pc$_2$Y]$^-$·TBA$^+$ with the ratio of [Ln]/[Y] = 1/49 in acetone. All measurements were performed using the micro-SQUID technique (SQUID is superconducting quantum interference device).[21] The field was aligned parallel to the easy axis of magnetization using the transverse field method.[22]

Figure 1 shows magnetization versus field measurements for the diluted [Pc$_2$Tb]$^-$ sample at 0.04 K and for several field scan rates. These hysteresis loops present clear staircase-like structure indicating the occurrence of QTM. Such measurements were also performed on less diluted and undiluted [Pc$_2$Tb]$^-$·TBA$^+$ samples revealing similar features that were however significantly broadened because of magnetic dipolar interactions among adjacent molecules. It should be noted that the figure shows no clear step at $H = 0$.

In known transition-metal-cluster SMMs, where energy separations between substates with different $|S_z|$ values are of the order of 1 to 10 cm$^{-1}$, QTM occurs when energy levels of two substates coincide under an appropriate magnetic field and the states are brought to resonance. In the lanthanide single-ion SMM, [Pc$_2$Tb]$^-$, such level crossings occur only at very high fields because the substates are separated by a few hundred cm$^{-1}$.[20,23,24] This situation is illustrated in Figure 2-a, a Zeeman diagram for the $J = 6$ ground multiplet of [Pc$_2$Tb]$^-$·TBA$^+$ obtained using the ligand-field (LF) parameters previously determined.[19]

Figure 2-b shows the enlargement of the area around zero field for the lowest $J_z$ = ±6 substates. Under the assumption that all off-diagonal matrix elements of the LF Hamiltonian are negligible,[20] no mixing occurs among the two sublevels at $H = 0$ (solid lines). In case of $C_4$ symmetry LF terms $A_4^4\langle r^4\rangle O_4^4$ and $A_6^4\langle r^6\rangle O_6^4$, which comprise the off-diagonal elements that couple $|J_z\rangle$ and $|J_z - 4\rangle$ states, can take non-zero values.[25] The plot with dotted lines is the Zeeman diagram are calculated with a small non-zero $A_4^4\langle r^4\rangle$ value. This illustrates that the $J_z = ±6$ substates are brought to resonance at $H = 0$, giving rise to an "avoided level crossing" which allows QTM to occur.

This picture itself, however, provides only insufficient explanation for the step structures observed at non-zero magnetic fields. Terbium has a nucleus with $I = 3/2$ spin in a natural abundance of 100 %. It is therefore necessary to take into account the interaction between the $(4f)^8$ system and the nucleus. Exact numerical diagonalization of a [$(2J+1)(2I+1) \times (2J+1)(2I+1)$] matrix, including the above ligand-field parameters, the hyperfine interaction $A_{hf}\mathbf{J}\cdot\mathbf{I}$, and nuclear quadrupole interaction term $P\{I_z^2 - 1/3I(I+1)\}$ was performed. Figure 3 shows the Zeeman diagram for the eight $|J_z\rangle|I_z\rangle$ states created from the combinations of the $J_z = ±6$ doublets and $I = 3/2$ quartets. The level intersections are seen at 13 magnetic field positions. Using $A_{hf} = 0.0173$ cm$^{-1}$ and $P = 0.010$ cm$^{-1}$, all step positions observed in the Figure 1 are reproduced. The four largest avoided level crossings are indicated. An important conclusion drawn from this numerical study is that the inclusion of the nuclear quadrupole term is mandatory to explain the seemingly irregularly arranged staircase structures in the hysteresis loops of [Pc$_2$Tb]$^-$. The hyperfine interaction term alone cannot account for all the steps observed. This is in sharp contrast to the equidistantly positioned steps observed for Ho ions in LiYF$_4$ matrix, which have been fully explained by the hyperfine interaction alone.[17]

Figure 4 shows hysteresis loops for the [Pc$_2$Dy]$^-$ complex at 0.04 K with varied field scan rates. A steep drop of magnetization is observed near zero field, unlike the Tb case. Apart this zero field step, only slight indications of steps can be seen at about $H = 7$ mT and 14 mT. Small remanent magnetization is present at low sweep rates.

In contrast to the Tb complex, the situation of Dy is more complicated. There are seven naturally occurring isotopes $^{156}$Dy, $^{158}$Dy, $^{160}$Dy, $^{161}$Dy, $^{162}$Dy, $^{163}$Dy and $^{164}$Dy with a natural abundance of 0.06, 0.01, 2.34, 18.91, 25.51, 24.90 and 28.18 %, respectively. $^{161}$Dy and $^{163}$Dy have a nuclear spin of $I = 5/2$ while other nuclear species are $I = 0$.

In the cases of $I = 0$, there is only one level crossing at $H = 0$. However, no tunneling should occur because of the Kramers theorem of spin parity. This theorem asserts that no matter how unsymmetric the crystal field, an ion possessing an odd number of electrons (that is a half-integer spin system) must have a ground state that is at least doubly degenerate (that is no tunnel splitting), even in the presence of crystal fields and spin-orbit interactions.[26] Hence, the Dy complexes with $I = 0$ should not contribute to the step structure. On the other hand, the Dy complexes with a nuclear spin $I = 5/2$ have avoided level crossings because two coupled half-integer spins ($J = 13/2$ and $I = 5/2$) lead to an integer total spin. Figure 5 shows the Zeeman diagram for the lowest sublevels calculated for the $I = 5/2$ case using the same value for $A_4^4\langle r^4 \rangle$ as in the above Tb case and the LF parameters determined previously.[20] Five avoided level crossings are seen. The hyperfine constant $A_{hf}$ was assumed to be 0.0042 cm$^{-1}$ in order to coincide the avoided level crossings with the positions indicated by the broken lines in Figure 4. A major difference from the Tb case concerns the tunnel splitting: the gaps are smaller than the corresponding values of the Tb complex by 2 orders of magnitude. This may be the reason why no clear step structures are observed at $H \neq 0$. Generally, a greatly reduced gap leads to significantly smaller tunneling probability according to the Landau-Zener model.[27]

In conclusion, we have shown for the first time the occurrence of QTM in the first lanthanide SMMs, [Pc$_2$Tb]$^-$ and [Pc$_2$Dy]$^-$, by hysteresis loops measurements of doped single crystals in the subkelvin temperature range. There is a fundamental difference in the mechanism of QTM between the lanthanide SMMs and the previously known transition-metal-cluster SMMs. The latter exhibits QTM between different substates $|S_z|$ whereas the former manifests QTM between the entangled states $|J_z\rangle|I_z\rangle$ of the electron and nuclear spin system.[17] The Tb complex with $J_z = \pm 6$ ground doublet and $I = 3/2$ nuclear spin gives avoided level crossings at the intersection of $|6\rangle|I_z\rangle$ and $|-6\rangle|I_z\rangle$. In the Dy case, only $^{161}$Dy and $^{163}$Dy complexes, having non-zero nuclear spin of $I = 5/2$, can exhibit avoided level crossings at the intersection of $|13/2\rangle|I_z\rangle$ and $|-13/2\rangle|I_z'\rangle$.

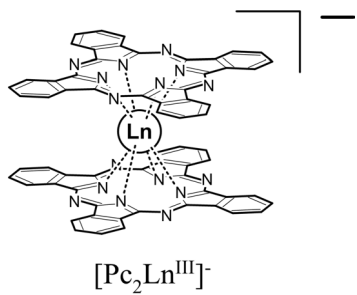

[Pc$_2$Ln$^{III}$]$^-$

**Scheme 1.**   [Pc$_2$Ln]$^-$   (Ln = Y, Tb or Dy)

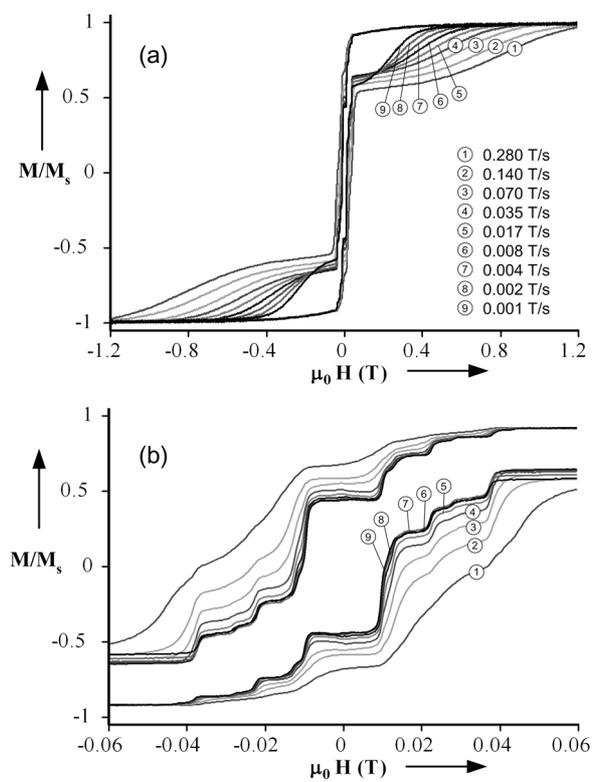

**Figure 1**. (a) Hysteresis loops at 0.04 K for a single crystal of [(Pc)$_2$Tb$_{0.02}$Y$_{0.98}$]$^-$·TBA$^+$ measured at several field scan rates. The applied magnetic field was aligned along the easy axis of magnetization. (b) Enlargement of the hysteresis loops in (a).

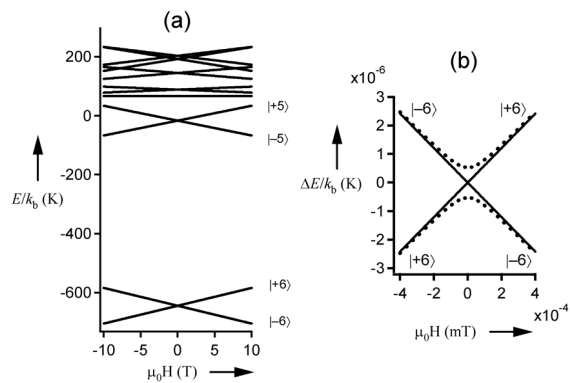

**Figure 2.** (a) Zeeman energy diagrams as a function of longitudinal magnetic field for the $J = 6$ ground multiplet with the ligand field parameters determined for [Pc$_2$Tb]$^-$·TBA$^+$ in a previous work:[19] $A_2^0\langle r^2\rangle$ = 414 cm$^{-1}$, $A_4^0\langle r^4\rangle$ = –228 cm$^{-1}$ and $A_6^0\langle r^6\rangle$ = 33 cm$^{-1}$. (b) Enlargement of the region around the intersection between the lowest substates. Dotted lines are obtained with an additional $A_4^4\langle r^4\rangle O_4^4$ ligand field term with $A_4^4\langle r^4\rangle$ = 10 cm$^{-1}$. The plot shows a tunnel splitting of $1.05 \times 10^{-6}$ K.

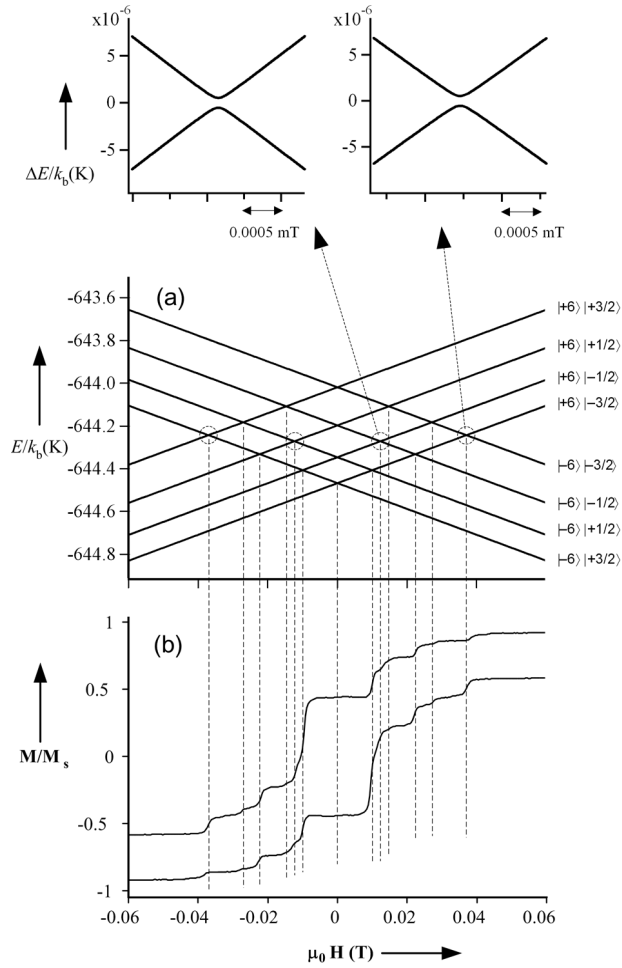

**Figure 3**. (a) Zeeman diagrams for the lowest $J_z = \pm 6$ substates with $I = 3/2$ nucleus calculated with the LF parameters used in the Figure 2-b, $A_{hf} = 0.0173$ cm$^{-1}$ for the hyperfine interaction term $A_{hf}\mathbf{J}\cdot\mathbf{I}$ and $P = 0.010$ cm$^{-1}$ for the nuclear quadrupole interaction term $P\{I_z^2 - 1/3I(I+1)\}$. The circles indicate the positions where avoided level crossing occurs. (b) Hysteresis loop at 0.04 K for a single crystal of [(Pc)$_2$Tb$_{0.02}$Y$_{0.98}$]$^-$·TBA$^+$ measured at 0.001 T/s.

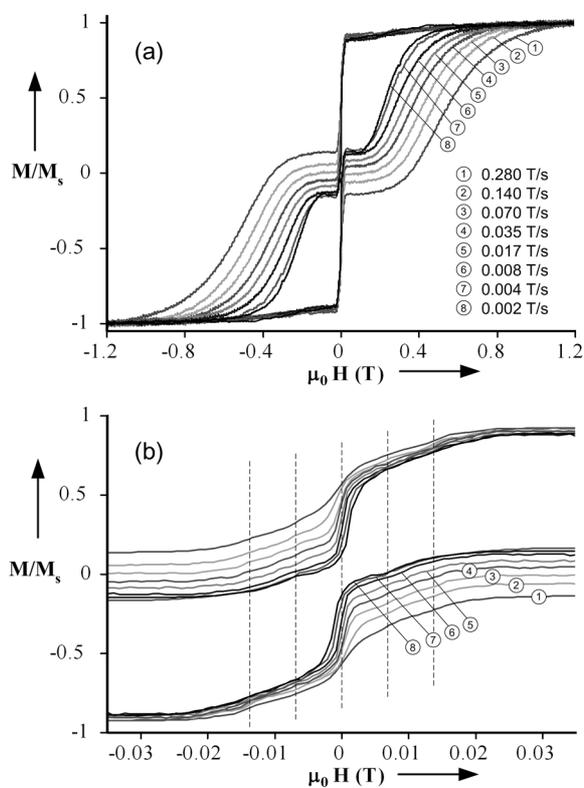

**Figure 4**. (a) Hysteresis loops at 0.04 K for a single crystalline sample of $[(Pc)_2Dy_{0.02}Y_{0.98}]^- \cdot TBA^+$ measured at several field scan rates. The applied magnetic field was aligned along the easy axis of magnetization. (b) Enlargement of the hysteresis loops in (a). The broken lines indicate the expected quantum resonances.

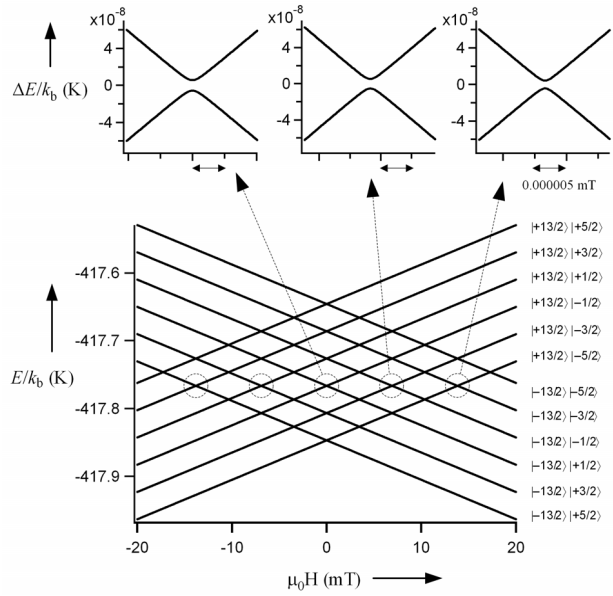

**Figure 5**. Energy diagrams in the function of longitudinal magnetic field for the lowest $J_z = \pm 13/2$ substates with $I = 5/2$ nucleus calculated with the LF parameters previously determined [19] and $A_4^4 \langle r^4 \rangle = 10$ cm$^{-1}$. The hyperfine constant is assumed to be $A_{hf} = 0.0042$ cm$^{-1}$ for the interaction term $A_{hf}\mathbf{J}\cdot\mathbf{I}$. The circles indicate the positions where avoided level crossing occurs.